\documentclass[prb,twocolumn,aps,showpacs]{revtex4}
\usepackage{graphicx}%
\usepackage{dcolumn}
\usepackage{amsmath}
\usepackage{amssymb}
\usepackage{epsfig}
\usepackage[latin1]{inputenc}
\usepackage{amsfonts}
\include{biblio}
\usepackage{color}
\usepackage{needspace}
\usepackage{makeidx}
\begin{document}
\title{Collective modes of a spin-orbit-coupled superfluid Fermi gas  in a two-dimensional optical lattice: a comparison between the Gaussian approximation and the Bethe-Salpeter approach}
\author{Zlatko  Koinov, Rafael Mendoza }\affiliation{Department of Physics and Astronomy,
University of Texas at San Antonio, San Antonio, TX 78249, USA}
\email{Zlatko.Koinov@utsa.edu}
 \begin{abstract}
A functional integral technique and a Legendre
transform are used to give a systematic derivation of the Schwinger-Dyson equations  for the generalized  single-particle Green's function and the Bethe-Salpeter equation for the two-particle Green's function and the associated collective modes of a population-imbalanced spin-orbit-coupled atomic Fermi gas loaded in a two-dimensional optical lattice at zero temperature. The collective-mode excitation energy is calculated within the Gaussian approximation, and from the Bethe-Salpeter equation in the generalized random phase approximation assuming the existence of a Sarma superfluid state. It is found that the Gaussian approximation  overestimates the speed of sound of the Goldstone mode. More interestingly, the Gaussian approximation  fails to reproduced the rotonlike structure of the collective-mode dispersion which appears after the linear part of the dispersion in the Bethe-Salpeter approach.

We use the Gaussian approximation and the Bethe-Salpeter approach to investigate the speed of sound of a balanced spin-orbit-coupled atomic Fermi gas near the boundary of the topological phase transition driven by an out-of-plane Zeeman field. It is shown that within both approaches,  the minimum of the speed of sound  is located at the topological phase transition boundary, and this fact  can be used to  confirm the existence of a topological phase transition.
\end{abstract}\pacs{03.75.Kk, 03.75.Ss, 67.85.Lm}
 \maketitle
\section{Introduction}
Topological superfluidity is an interesting state of matter, partly because it is
associated with quasiparticle excitations which are Majorana fermions.\cite{M} The basic physics behind the emergence of the Majorana fermion
excitations is the existence of s-wave superfluidity, nonvanishing  spin-orbit coupling (SOC), and Zeeman splitting. In this context, calculating the dispersion of the collective modes when the pseudospin of atoms can couple with not only the effective Zeeman field, but also with the
 orbital degrees of freedom of atoms is an important and interesting problem by itself.

The recent experimental breakthroughs in realization of spin-orbit
coupling (SOC)\cite{SOC1,SOC2,SOC3,SOC4,SOC5,SOC6,SOC7,SOC8,SOC9,SOC10,SOC11} in ultracold atomic gases have opened up possibilities for investigating the phase diagram, the single-particle and  the collective-mode excitations of  superfluid Fermi gases  in  optical lattices in the presence of SOC and Zeeman fields. It is widely accepted among the optical-lattice community that the attractive Fermi-Hubbard model captures the s-wave superfluidity of cold fermion atoms in optical lattices. According to this model, two fermion atoms of opposite pseudospins on the same site have  an attractive interaction energy $U$, while the probability to tunnel to a neighboring site is given by the hopping parameter $J$:
\begin{equation}H_U=-\sum_{<i,j>,\sigma}J_{\sigma}\psi^\dag_{i,\sigma}\psi_{j,\sigma}
-U\sum_i \widehat{n}_{i,\uparrow}
\widehat{n}_{i,\downarrow}-\sum_{i,\sigma}\mu_\sigma\widehat{n}_{i,\sigma}.\nonumber\end{equation}
Here, $J_{\sigma}$ is  the tunneling strength of the atoms between
nearest-neighbor sites,  and
 $\widehat{n}_{i,\sigma}=\psi^\dag_{i,\sigma}\psi_{i,\sigma}$ is the
density operator on site $i$. The Fermi operator
$\psi^\dag_{i,\sigma}$ ($\psi_{i,\sigma}$) creates (destroys) a
fermion on the lattice site $i$  with pseudospin projection
$\sigma=\uparrow,\downarrow$. The symbol $\sum_{<ij>}$ means sum over nearest-neighbor
sites of the two-dimensional (2D) lattice.

In this paper we assume the existence of nonvanishing Rashba SOC in the xy plane and a
Zeeman field along the z direction, so the Hamiltonian of the system is $\widehat{H}=\widehat{H}_U+\widehat{H}_{SOC}+\widehat{H}_Z$. In the case of a 2D optical lattice the SOC part of the Hamiltonian is given by\cite{Ham1,Ham2}
\begin{equation}\widehat{H}_{SOC}=-\imath\lambda \sum_{<i,j>} \left(\psi^\dag_{i,\uparrow}, \psi^\dag_{j,\downarrow}\right)\left(\overrightarrow{\sigma}\times \textbf{d}_{i,j}\right)_z\left(%
\begin{array}{c}
\psi_{i,\uparrow}\\\psi_{j,\downarrow}\\
\end{array}%
\right),\nonumber\end{equation}
where $\lambda$ is the Rashba SO coupling coefficient, $\overrightarrow{\sigma}=(\sigma_x, \sigma_y,\sigma_z)$, $\sigma_{x,y,z}$ are the Pauli matrices, and $\textbf{d}_{i,j}$ is a unit vector along the
line that connects site $j$ to $i$.

The out-of-plane Zeeman field is described by the term $\widehat{H}_Z$:
\begin{equation}\widehat{H}_{Z}=h_z\sum_{i} \left(\psi^\dag_{i,\uparrow}, \psi^\dag_{i,\downarrow}\right)\sigma_z\left(%
\begin{array}{c}
\psi_{i,\uparrow}\\\psi_{i,\downarrow}\\
\end{array}%
\right).\nonumber\end{equation}

The simplest way to study the collective excitations of the above-mentioned  model is to apply functional integral technique which requires the representation of the Hubbard interaction in terms of squares of one-body charge and spin operators. It is possible to resolve the Hubbard interaction into quadratic forms of spin and electron number operators in an infinite number of ways by applying the Hubbard-Stratonovich transformation. If no approximations were made in evaluating the functional integrals, it would no matter which of the ways is chosen. When approximations are taken, the final result depends on a particular form chosen.

One of the most common ways to apply the Hubbard-Stratonovich transformation is to introduce the energy gap as an order parameter field, which allows us to integrate out the fermion fields and to arrive at an effective action. The next steps are to consider the  state which corresponds to the saddle point of the effective action, and to  write the effective action as a series in powers of the fluctuations and their derivatives. The exact result can be obtained by explicitly calculating the terms up to second order in the fluctuations and their derivatives. This approximation is called the Gaussian approximation.

In the case of vanishing SOC and $h_z=0$, the Gaussian approximation  provides the following equation for the collective mode dispersion $\omega(\textbf{Q})$ of homogeneous superfluid Fermi gases at a zero temperature:\cite{G1}
\begin{equation}0=1+U\left(I_{\gamma,\gamma}+I_{l,l}\right)+
U^2\left(I_{\gamma,\gamma}I_{l,l}-J^2_{\gamma,l}\right),
\label{Gauss}\end{equation}
where $I_{\gamma,\gamma}=I_{\gamma,\gamma}(\omega,\textbf{Q}), I_{l,l}=I_{l,}(\omega,\textbf{Q})$ and $J_{\gamma,l}=J_{\gamma,l}(\omega,\textbf{Q})$ are defined in the Appendix.

Instead of introducing the energy gap as an order parameter field, one can transform the quartic terms to a quadratic forms  by introducing  a four-component boson field which mediates the interaction of fermions.   This approach  is similar to the situation in quantum electrodynamics, where the photons mediate the interaction of electric charges.   This similarity allows us to apply  the powerful method of Legendre transforms, to derive  the Schwinger-Dyson\cite{Schwinger,Dyson} (SD) equation $G^{-1}=G^{(0)-1}-\Sigma$,  and the Bethe-Salpeter\cite{BetheS} (BS) equation $[K^{(0)-1}-I]\Psi=0$ for the poles of the single-particle Green's function, $G$, and the poles of the two-particle Green's function, $K$, respectively. Here,  $G^{(0)}$ is the
 free single-particle propagator, $\Sigma$ is the fermion self-energy,
$I$ is the BS kernel, and the two-particle free propagator $K^{(0)}
= GG $ is a product of two fully dressed single-particle Green's
functions. The kernel of the BS equation is defined as a sum of the direct interaction, $I_d=\delta\Sigma^F/ \delta G$, and the exchange interaction  $I_{exc}=\delta\Sigma^H/ \delta G$, where $\Sigma^F$ and $\Sigma^H$ are the Fock and the Hartree parts of the fermion self-energy $\Sigma$. Since the fermion self-energy
 depends on the two-particle Green's function, the
positions of both poles must to be obtained by solving
 the SD and BS equations self-consistently.

It is widely accepted that the generalized random phase approximation (GRPA) is a good approximation for the collective excitations in a weak-coupling regime, and therefore, it can be used to separate the solutions of the SD
and the BS equations. In this approximation, the single-particle excitations are obtained in the mean field approximation (or by solving the Bogoliubov-de Gennes equations in a self-consistent fashion); while the collective modes are
obtained by solving the BS equation in which the single-particle
Green's functions are calculated in Hartree-Fock  approximation, and the BS
kernel is obtained by summing ladder and bubble diagrams. The BS equation in the GRPA has been used to obtain the collective-mode spectrum
of an imbalanced Fermi gas in a deep optical lattice.\cite{KMF,RM,KPM,ZGK}

In the GRPA, the equation (\ref{Gauss}) for the collective mode dispersion $\omega(\textbf{Q})$ of homogeneous superfluid Fermi gases at a zero temperature is replaced by the following one:\cite{BR}
\begin{equation}\begin{split}&
0=1+\left(I_{l,l}+I_{\gamma,\gamma}+I_{m,m}\right)U+(I_{l,l}I_{\gamma,\gamma}-J^2_{\gamma,l}\\&-I^2_{l,m}+I_{l,l}I_{m,m}-J^2_{\gamma,m}
+I_{m,m}I_{\gamma,\gamma})U^2+
(-I_{m,m}J^2_{\gamma,l}\\&+2I_{l,m}J_{\gamma,l}J_{\gamma,m}
-I_{l,l}J^2_{\gamma,m}-I_{\gamma,\gamma}I^2_{l,m}+I_{l,l}I_{m,m}I_{\gamma,\gamma})U^3.
\label{F1}\end{split}\end{equation}

\begin{figure}
\includegraphics[scale=0.7]{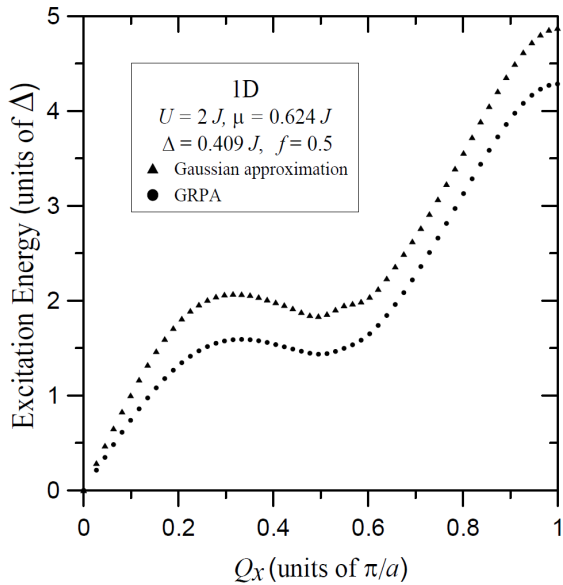}
    \caption{Collective mode dispersions of the Goldstone mode in a one-dimensional optical lattice in the weak coupling limit obtained by applying the Gaussian approximation (triangles), and by the GRPA (circles).   The lattice constant is $a$.
We set the filling factors to be $f_\uparrow=f_\downarrow=0.25$, and the strength of the interaction is $U=2J$. The mean field
superfluid gap and the chemical potential are
$\Delta=0.409J$ and $\mu=0.624J$.}\label{Fig1}\end{figure}

\begin{figure}
\includegraphics[scale=0.85]{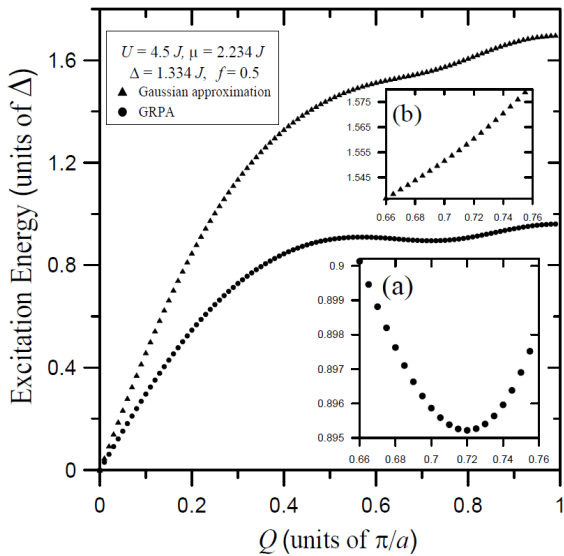}
    \caption{Collective mode dispersions $\omega(Q_x,Q_y)$ of the Goldstone mode in a two-dimensional optical lattice along the $Q_x=Q_y=Q$ direction in the weak coupling limit obtained by applying the Gaussian approximation (triangles), and by the GRPA (circles).  The rotonlike
structure is seen in the GRPA spectrum (inset (a)), while the roton minimum does not appear in the Gaussian approximation (inset (b)).
The filling factors, and the strength of the interaction are $f_\uparrow=f_\downarrow=0.25$ and $U=4.5J$. The mean field
superfluid gap and the chemical potential are
$\Delta=1.334J$ and $\mu=2.234J$.}\label{Fig2}\end{figure}

In Fig. \ref{Fig1} and Fig. \ref{Fig2}, we  present the collective-mode dispersions $\omega(Q_x)$ in a one-dimensional (1D) optical lattice, and $\omega(Q_x,Q_y)$ in a 2D  lattice along the $Q_x=Q_y=Q$ direction, both obtained by numerically solving Eqs. (\ref{Gauss}) and (\ref{F1}).  The speed of sound is defined by the slope of the Goldstone-mode dispersion in the limit $Q\rightarrow 0$. When the Gaussian approach is used, the speed of sound in the 1D case is  $u=1.30 Ja/\hbar$, while the GRPA provides $u=0.96 Ja/\hbar$. In the 2D case, the corresponding speeds are  $u=2.00 Ja/\hbar$, and $u=1.30 Ja/\hbar$, respectively. In both cases, the speed of sound is overestimated by the Gaussian approximation. In a diagrammatic language,  Eq. (\ref{Gauss}) can be derived by summation of the infinite sequences of graphs in the ladder approximation.\cite{CKS} Thus, one can say that the speed of sound  decreases when the exchange interaction, represented by bubble diagrams, is taken into account. Another interesting fact is that in 1D case both approximations provide the existence of a roton minimum, while the inset (b) in the Fig. \ref{Fig2} shows no roton minimum within the Gaussian approximation.

In what follows, we study the collective-mode dispersion of species of Fermi atoms with equal, or imbalance,  population in two pseudospin states loaded in a 2D optical lattice in the presence of both Zeeman field and nonvanishing Rashba type of spin-orbit coupling.  To the best of our knowledge, the Gaussian approximation is the only approximation that has been used to obtain the speed of sound   in the presence of  the SOC and the  Zeeman field effects.\cite{GSOC1,GSOC2,GSOC3,GSOC4,GSOC5,GSOC6,GSOC7,GSOC8}  In a view of the facts that: (i) in a lattice system  the bubble diagrams induce the instability due to the charge-density-wave (CDW) fluctuations, (ii) because of the strong CDW fluctuations  the collective-mode spectrum has a characteristic roton-like structure which lies below the particle-hole continuum (for a detailed discussion of the effects of the
CDW fluctuations on the stability of the collective modes, see Ref. [\onlinecite{Jap}]), and (iii) the speed of sound is overestimated in the Gaussian approximation,  one may well ask what would be the difference between the collective-mode dispersion and the corresponding speed of sound in the presence of the SOC and Zeeman fields when calculated within the Gaussian approximation, and from the BS formalism by summing infinite sequences of ladder and bubble diagrams.

The main goal of the present study is to obtain the BS equation  for the collective modes in the GRPA which takes into account the SOC and the Zeeman effects. In Sec. II, we derive the BS equation for the poles of the two-particle Green's function, which allows us to obtain numerically  the dispersion of the collective modes in the presence of both the Zeeman field and the Rashba types of SOC. In Sec. III,  the roton-like structure of the collective-mode spectrum of an imbalanced 2D  Fermi gas in the presence of SOC is obtained by solving the BS equation in the GRPA. It turns out that the Gaussian approximation and the BS approach, both provide a very similar slope of the linear part of the Goldstone dispersion, but there is no roton minimum within the Gaussian approximation. In the same Section, we study the speed of sound as a function of the strength of the Zeeman field near the  topological quantum phase transition boundary from a nontopological superfluid state with a fully gapped fermionic spectrum to a topological superfluid state.
 \section{The Bethe-Salpeter equation}
\subsection{The functional-integral formulation of the Hubbard
model}
We consider an imbalanced mixture of an atomic Fermi gas
of two hyperfine states (described by pseudospins $\sigma=\uparrow,\downarrow$) with contact interaction
loaded into a 2D square optical lattice.  In the imbalanced case, there are different amounts of $M_\uparrow$ and $M_\downarrow$ atoms in each  state achieved by considering different chemical potentials $\mu_\uparrow$ and $\mu_\downarrow$. The total number of atoms $M = M_\uparrow+M_\downarrow$ is distributed along $N$ sites, and the corresponding filling
factors $f_{\uparrow,\downarrow}=M_{\uparrow,\downarrow}/N$ are smaller than unity.

From a theoretical point of view, the corresponding expressions for the Green's functions cannot be
evaluated exactly because the  interaction part
of the Hubbard Hamiltonian is quartic in the fermion fields. The simplest way to solve this problem is to
apply the so-called mean-field decoupling of the quartic
interaction.\cite{Ham1,Ham2} To go beyond the mean-field approximation, we apply the
idea that one can transform the quartic term
 into  quadratic form by making the
Hubbard-Stratonovich  transformation  for the fermion operators. In
contrast to the previous approaches, wherein  after performing the
Hubbard-Stratonovich  transformation the fermion degrees of freedom
are integrated out; we decouple the quartic problem by introducing a
model system which consists of a multi-component boson field
$A_\alpha$ interacting with  fermion fields $\psi^\dagger$ and
$\psi$.

 The functional-integral formulation of the Hubbard
model requires the representation of the Hubbard interaction $\widehat{H}_U$
 in terms of squares of one-body charge and spin operators.
It is known that when approximations are made, the
final result depends on the particular form chosen. Thus, one should
check that the results obtained with a certain Hubbard-Stratonovich
transformation are consistent with the results obtained with
canonical mean-field approximation. It can be seen that our approach
to the Hubbard-Stratonovich transformation provides  results
consistent with the results obtained with the mean-field
approximation, i.e. one can derive the mean-field gap equation using
the collective-mode dispersion $\omega(Q)$ in the limit
$Q\rightarrow 0$ and $\omega\rightarrow 0$.

 The Green's functions in the  functional-integral approach are defined by means
  of the so-called generating functional with sources for the boson and fermion  fields, but
    the corresponding functional integrals  cannot be evaluated
exactly because the interaction part of the Hubbard Hamiltonian  is
quartic in the Grassmann fermion fields. However, we can transform
the quartic terms to a quadratic form by introducing  a model system
which consists of a four-component boson field $A_{\alpha}(z)$ ($\alpha=1,2,3,4,\quad  z=(\textbf{r}_i,v),\quad 0\leq v \leq \beta=(k_BT)^{-1}$, $k_B$ is the Boltzmann constant)
 interacting with fermion fields
$\widehat{\overline{\psi}} (y)=\widehat{\Psi}^\dag (y)/\sqrt{2}$ and
$\widehat{\psi}(x)=\widehat{\Psi}(x)/\sqrt{2}$, where
\begin{equation}\widehat{\Psi}(x)=\left(%
\begin{array}{c}
  \psi_\uparrow(x) \\
  \psi_\downarrow(x) \\
  \psi^\dag_\uparrow(x)\\
  \psi^\dag_\downarrow(x)\\
\end{array}%
\right),\quad \widehat{\Psi}^\dag
(y)=\left(\psi^\dag_\uparrow(y)\psi^\dag_\downarrow(y)
\psi_\uparrow(y)\psi_\downarrow(y)
\right),\label{NG1}\end{equation}

The field operators (\ref{NG1}) allow us to define the  generalized
single-particle Green's function by using a tensor product of these two matrices. The corresponding Green's function, represented by a  $4\times 4$ matrix,  includes all possible
thermodynamic averages:
\begin{equation}\widehat{G}(x_1;y_2)=-<\widehat{T}_u\left(\widehat{\Psi}(x_1)\otimes\widehat{\overline{\Psi}}(y_2)\right)>
. \label{EGF}
\end{equation}

The action of the above-mentioned model system is assumed to
be of the following form
 $S= S^{(F)}_0+S^{(B)}_0+S^{(F-B)}$, where $S^{(F)}_0=\widehat{\overline{\psi}
}(y)\widehat{G}^{(0)-1}(y;x)\widehat{\psi} (x)$, $
S^{(B)}_0=\frac{1}{2}A_{\alpha}(z)D^{(0)-1}_{\alpha
\beta}(z,z')A_{\beta}(z')$, $ S^{(F-B)}=\widehat{\overline{\psi}}
(y)\widehat{\Gamma}^{(0)}_{\alpha}(y,x\mid z)\widehat{\psi}
(x)A_{\alpha}(z)$.  The action $S^{(F)}_0$ describes the fermion  part of the system.
 The generalized inverse Green's function
 of free fermions
$\widehat{G}^{(0)-1}(y;x)$ is given by the following  $4\times 4$ matrix:
\begin{equation}\begin{split}&\widehat{G}^{(0)-1}(y;x)=\\&\sum_{\textbf{k},\omega_m}\exp\left[\imath
\textbf{k.}(\textbf{r}_i-\textbf{r}_{i'})-\omega_m(u-u')\right]
G_{n_1n_2}^{(0)-1}(\textbf{k},\imath\omega_m) , \nonumber
\end{split}\end{equation}
 where  $G_{11}^{(0)-1}(\textbf{k},\imath\omega_m)=
 -G_{33}^{(0)-1}(-\textbf{k},-\imath\omega_m)
 $, and
$-G_{22}^{(0)-1}(-\textbf{k},-\imath\omega_m)=G_{44}^{(0)-1}(\textbf{k},\imath\omega_m)$.
The symbol $\sum_{\omega_m}$ is used to denote $\beta^{-1}\sum_{m}$
(for fermion fields $\omega_m=
   (2\pi/\beta)(m +1/2) ;m=0, \pm 1, \pm 2,... $).
   In the case of the  population-imbalanced Fermi gas with a Rashba SO coupling and an out-of-plane Zeeman field,  the non-interacting Green's
   function is:\cite{GSOC6}\begin{widetext}
     \begin{equation}
 \widehat{G}^{(0)-1}(\textbf{k},\imath\omega_m)=\left(%
\begin{array}{cccc}
  \imath\omega_m-\xi_\uparrow(\textbf{k})-h_z&2\lambda\left(\sin k_x+\imath\sin k_y\right)&0&0  \\
  2\lambda\left(\sin k_x-\imath\sin k_y\right)&\imath\omega_m-\xi_\downarrow(\textbf{k})+h_z &0&0  \\
 0 &0 &\imath\omega_m+\xi_\uparrow(\textbf{k})+h_z&-2\lambda\left(\sin k_x-\imath\sin k_y\right)\\
 0 &0 & 2\lambda\left(\sin k_x+\imath\sin k_y\right)&\imath\omega_m+\xi_\downarrow(\textbf{k})-h_z
\end{array}%
\right).
 \label{GF0}\end{equation}
Here, $\xi_{\uparrow,\downarrow}(\textbf{k})=2J_{\uparrow,\downarrow}(1-\cos k_x)+2J_{\uparrow,\downarrow}(1-\cos k_y)-\mu_{\uparrow,\downarrow}$ is the tight binding form of the electron energy (the lattice constant $a=1$).

 The action $S^{(B)}_0$ describes
the boson field which mediates
the  fermion-fermion on-site interaction in the Hubbard Hamiltonian. The bare boson propagator in $S^{(B)}_0$  is defined as:
\begin{equation}\widehat{D}^{(0)}
(z,z')=\delta(v-v')U\delta{j,j'}\left(%
\begin{array}{cccc}
  0&1&0&0  \\
 1 &0 &0&0  \\
 0 &0 &0&0\\
 0 &0 & 0&0
\end{array}%
\right).\nonumber\end{equation}  The Fourier transform of this boson
propagator is given by
\begin{equation}
\widehat{D}^{(0)} (z,z')=\frac{1}{N} \sum_\textbf{k}\sum_{\omega_p}
 e^{\left\{\imath\left[\textbf{k.}\left(\textbf{r}_j-\textbf{r}_{j'}\right)
 -\omega_p\left(v-v'\right)\right]\right\}}\widehat{D}^{(0)}(\textbf{k}),
\widehat{D}^{(0)}(\textbf{k})= \left(%
\begin{array}{cccc}
  0&U &0&0  \\
 U  &0 &0&0  \\
 0 &0 &0&0 \\
 0 &0 & 0&0
\end{array}%
\right).\label{FTD0}\end{equation}

 The interaction between the fermion and the boson
 fields is described by the action $S^{(F-B)}$.
 The bare vertex
$\widehat{\Gamma}^{(0)}_{\alpha}(y_1;x_2\mid
z)=\widehat{\Gamma}^{(0)}_{\alpha}(i_1,u_1;i_2, u_2\mid
j,v)=\delta(u_1-u_2)\delta(u_1-v)\delta_{i_1i_2}\delta_{i_1j}\widehat{\Gamma}^{(0)}(\alpha)$
is a $4\times 4$ matrix, where
\begin{equation}
\widehat{\Gamma}^{(0)}(\alpha)=\frac{1}{2}(\gamma_0+\alpha_z)\delta_{\alpha1}
+\frac{1}{2}(\gamma_0-\alpha_z)\delta_{\alpha2}+
\frac{1}{2}(\alpha_x+\imath\alpha_y)\delta_{\alpha3}+
\frac{1}{2}(\alpha_x-\imath\alpha_y)\delta_{\alpha4}.\label{Gamma0}\end{equation}
The Dirac matrix $\gamma_0$  and the  matrices
 $\widehat{\alpha}_i$  are defined as (when a four-dimensional space is used, the electron spin operators $\sigma_i$ has to be replaced by $\widehat{\alpha}_i  \gamma_0$ ):
$$\gamma_0=\left(%
\begin{array}{cccc}
  1&0&0&0  \\
 0&1&0&0  \\
 0& 0& -1&0  \\
 0& 0& 0&-1  \\
\end{array}%
\right),\quad  \widehat{\alpha}_i=\left(%
\begin{array}{cc}
  \sigma_i & 0  \\
 0& \sigma_y\sigma_i\sigma_y \\
\end{array}%
\right), i=x,y,z.$$

The relation between the Hubbard  model and our model system can be
demonstrated by applying the Hubbard-Stratonovich transformation for
the fermion operators:
\begin{equation}\int D\mu[A]\exp\left[\widehat{\overline{\psi}}
(y)\widehat{\Gamma}^{(0)}_{\alpha}(y;x|z)\widehat{\psi}(x)A_{\alpha}(z)\right]
=\exp\left[-\frac{1}{2}\widehat{\overline{\psi}}
(y)\widehat{\Gamma}^{(0)}_{\alpha}(y;x|z)\widehat{\psi}(x)
D_{\alpha,\beta}^{(0)}(z,z') \widehat{\overline{\psi}}
(y')\widehat{\Gamma}^{(0)}_{\beta}(y';x'|z')\widehat{\psi}(x')\right].\nonumber\end{equation}
 The functional measure $D\mu[A]$ is chosen to be:
$$
D\mu[A]=DAe^{-\frac{1}{2}A_{\alpha}(z)D_{\alpha,\beta}^{(0)-1}(z,z')
A_{\beta}(z')},\int \mu[A] =1.$$

 According to the field-theoretical approach, the expectation value of a general operator
$\widehat{O}(u)$ can be expressed as a functional integral over the
boson field $A$ and the Grassmann fermion fields
$\widehat{\overline{\psi}}$ and $\widehat{\psi}$:
\begin{equation}<\widehat{T}_u(\widehat{O}(u))>=\frac{1}{Z[J,M]}\int
D\mu[\widehat{\overline{\psi}},\widehat{\psi},A]\widehat{O}(u)
\exp\left[J_{\alpha}(z)A_{\alpha}(z)-M(\widehat{\overline{\psi}},\widehat{\psi})\right]|_{J=M=0},\label{Ex}\end{equation}
where the symbol $<...>$ means that the thermodynamic average is
made. The
functional $Z[J,M]$ is defined by
\begin{equation}
Z[J,M]=\int
D\mu[\widehat{\overline{\psi}},\widehat{\psi},A]
\exp\left[J_{\alpha}(z)A_{\alpha}(z)-M(\widehat{\overline{\psi}},\widehat{\psi})\right],\label{GW}
\end{equation}where the functional measure
$D\mu[\widehat{\overline{\psi}},\widehat{\psi},A]=DAD\widehat{\overline{\psi}}D\widehat{\psi}
\exp\left(S\right)$ satisfies the condition $\int
D\mu[\widehat{\overline{\psi}},\widehat{\psi},A]=1$. The quantity
$J_\alpha(z)$ is the source of the boson field, while the sources
$M_{ij}(y;x)$ of the fermion fields are included in the
$M(\widehat{\overline{\psi}},\widehat{\psi})$ term :
\begin{eqnarray}& M(\widehat{\overline{\psi}},\widehat{\psi})=
\psi^\dag_\uparrow(y)
M_{11}(y;x)\psi_\uparrow(x)+\psi^\dag_\downarrow(y)
M_{21}(y;x)\psi_\uparrow(x)+\psi^\dag_\uparrow(y) M_{12}(y;
x)\psi_\downarrow(x)\nonumber\\&+\psi^\dag_\downarrow(y)
M_{22}(y;x)\psi_\downarrow(x)+ \psi_\uparrow(y)
M_{31}(y;x)\psi_\uparrow(x)+\psi_\downarrow(y)
M_{41}(y;x)\psi_\uparrow(x)\nonumber\\&+\psi_\uparrow(y)
M_{32}(y;x)\psi_\downarrow(x)+\psi_\downarrow(y)
M_{42}(y;x)\psi_\downarrow(x)\nonumber\\& + \psi^\dag_\uparrow(y)
M_{13}(y;x) x)\psi^\dag_\uparrow(x)+\psi^\dag_\downarrow(y)
M_{23}(y;x)\psi^\dag_\uparrow(x)+\psi^\dag_\uparrow(y)
M_{14}(y;x)\psi^\dag_\downarrow(x)\nonumber\\&+\psi^\dag_\downarrow(y)
M_{24}(y;x)\psi^\dag_\downarrow(x) +\psi_\uparrow(y)
M_{33}(y;x)\psi^\dag_\uparrow(x)+\psi_\downarrow(y)
M_{43}(y;x)\psi^\dag_\uparrow(x)\nonumber\\&+\psi_\uparrow(y)
M_{34}(y;x)\psi^\dag_\downarrow(x)+\psi_\downarrow(y)
M_{44}(y;x)\psi^\dag_\downarrow(x).
 \label{M1}\end{eqnarray}

We shall now use  a functional derivative $\delta / \delta M(2;1)=\delta / \delta M_{n_2,n_1}(y_2;x_1)$, where $1=\{n_1,x_1\}$ and  $2=\{n_2,y_2\}$ are complex indices;
depending on the spin degrees of freedom, there are sixteen possible
derivatives.  By means of the definition (\ref{Ex}), all Green's functions can be expressed  in terms of the functional derivatives with respect to the
  corresponding  sources
 of the
generating functional of the connected Green's functions $W[J,M]=\ln
Z[J,M]$. Thus, we define the following Green's and vertex functions
which will be used to  analyze the collective modes of our model:

  The Boson Green's
function is $D_{\alpha \beta}(z,z')$ is a $4\times 4$ matrix defined
as $$D_{\alpha \beta}(z,z')=-\frac{\delta^2W}{\delta
J_{\alpha}(z)\delta J_{\beta}(z')}.$$

The generalized single-fermion Green's function $G_{n_1n_2}(x_1;y_2)$ is the
 matrix defined by Eq. (\ref{EGF}) whose elements are
$$G_{n_1n_2}(x_1;y_2)=-\delta W/\delta M_{n_2n_1}(y_2;x_1).$$
 Depending on the two spin degrees of freedom,
$\uparrow$ and $\downarrow$, there exist eight "normal" Green's
functions  and eight "anomalous" Green's functions. We introduce
Fourier transforms of the "normal"
$G_{\sigma_1,\sigma_2}(\textbf{k},u_1-u_2)=
-<\widehat{T}_u(\psi_{\sigma_1,\textbf{k}}(u_1)\psi^\dag_{\sigma_2,\textbf{k}}(u_2))>$,
and "anomalous"
$F_{\sigma_1,\sigma_2}(\textbf{k},u_1-u_2)=-<\widehat{T}_u(\psi_{\sigma_1,\textbf{k}}(u_1)
\psi_{\sigma_2,-\textbf{k}}(u_2))>$  one-particle Green's functions,
where $\{\sigma_1,\sigma_2\}=\uparrow,\downarrow$. Here
$\psi^+_{\uparrow,\textbf{k}}(u),\psi_{\uparrow,\textbf{k}}(u)$ and
$\psi^+_{\downarrow,\textbf{k}}(u),\psi_{\downarrow,\textbf{k}}(u)$
are the creation-annihilation Heisenberg operators. The Fourier
transform of the generalized single-particle Green's function is given by
\begin{eqnarray}&
\widehat{G}(1;2)=
\frac{1}{N}\sum_{\textbf{k}}\sum_{\omega_{m}}\exp\{\imath\left[\textbf{k.}\left(
\textbf{r}_{i_1}-\textbf{r}_{i_2}\right)-\omega_{m}(u_1-u_2)\right]\}
\left(%
\begin{array}{cc}
  \widehat{G}(\textbf{k},\imath\omega_m) & \widehat{F}(\textbf{k},\imath\omega_m)  \\
 \widehat{F}^\dag(\textbf{k},\imath\omega_m) & -\widehat{G}(-\textbf{k},-\imath\omega_m) \\
\end{array}%
\right).\label{FTGF}\end{eqnarray} Here, $\widehat{G}$ and
$\widehat{F}$ are $2\times 2$ matrices whose elements are
$G_{\sigma_1,\sigma_2}$ and $F_{\sigma_1,\sigma_2}$, respectively.

The two-particle
Green's function $K\left(%
\begin{array}{cc}
  n_1,x_1 & n_3,y_3  \\
  n_2,y_2 & n_4,x_4 \\
\end{array}%
\right)$ is defined as
\begin{equation}
K\left(%
\begin{array}{cc}
  n_1,x_1 & n_3,y_3  \\
  n_2,y_2 & n_4,x_4 \\
\end{array}%
\right)=K\left(%
\begin{array}{cc}
  1 & 3  \\
  2 & 4 \\
\end{array}%
\right)=\frac{\delta^2 W}{\delta M_{n_2n_1}(y_2;x_1)\delta
M_{n_3n_4}(y_3;x_4)}=-\frac{\delta G_{n_1n_2}(x_1;y_2)}{\delta
M_{n_3n_4}(y_3;x_4)} . \nonumber
\end{equation}

 The vertex function $\widehat{\Gamma}_{\alpha}(2;1 \mid
  z)$ for a given $\alpha$  is a $4 \times 4$ matrix whose elements are:
$$
\widehat{\Gamma}_{\alpha}(i_2,u_2;i_1,u_1 \mid
v,j)_{n_2n_1}=-\frac{\delta G_{n_2n_1}^{-1}(i_2,u_2;i_1,u_1)}{\delta
J_{\beta}(z')} D^{-1}_{\beta \alpha}(z',z). $$

Next, we shall obtain the corresponding equations of the boson and fermion Green's functions. The poles of these Green's functions provide the single-particle and the two-particle excitations.

It is known that the fermion self-energy (fermion mass
operator) $\widehat{\Sigma}(1;2)$ can be defined by means of the
 SD equations. They can be derived using the fact that the
measure $D\mu[\overline{\psi},\psi,A]$ is invariant under the
translations $\overline{\psi}\rightarrow
\overline{\psi}+\delta\overline{\psi}$ and  $A\rightarrow A+\delta
A$:
\begin{equation}
D^{(0)-1}_{\alpha
\beta}(z,z')R_\beta(z')+\frac{1}{2}Tr\left(\widehat{G}(1;2)\widehat{\Gamma}^{(0)}_{\alpha}(2;1\mid
z)\right)+J_\alpha(z)=0, \label{SE1}
\end{equation}
\begin{equation}
\widehat{G}^{-1}(1;2)-\widehat{G}^{(0)-1}(1;2)+\widehat{\Sigma}(1;2)+\widehat{M}(1;2)=0,
\label{SE2}
\end{equation}
where $R_\alpha(z)=\delta W/\delta J_\alpha(z)$ is the average boson
field. The fermion self-energy   $\widehat{\Sigma}$, is a $4\times
4$ matrix which can be written as a sum of Hartree
$\widehat{\Sigma}^H$ and Fock $\widehat{\Sigma}^F$ parts. The
Hartree part is a diagonal matrix whose elements are:
\begin{eqnarray}&
\Sigma^H(i_1,u_1;i_2,u_2)_{n_1n_2}=\frac{1}{2}
\widehat{\Gamma}_{\alpha}^{(0)}(i_1,u_1;i_2,u_2|j,v)_{n_1n_2}
D^{(0)}_{\alpha\beta}(j,v;j',v')\nonumber\\&\widehat{\Gamma}_{\beta}^{(0)}(i_3,u_3;i_4,u_4|j',v')_{n_3n_4}
G_{n_4n_3}(i_4,u_4;i_3,u_3) .\label{H1}
\end{eqnarray}

The Fock part of the fermion self-energy is given by:
\begin{eqnarray} &
\Sigma^F(i_1,u_1;i_2,u_2)_{n_1n_2}=-
\widehat{\Gamma}_{\alpha}^{(0)}(i_1,u_1;i_6,u_6|j,v)_{n_1n_6}
D^{(0)}_{\alpha\beta}(j,v;j',v')\nonumber\\&\widehat{\Gamma}_{\beta}^{(0)}(i_4,u_4;i_5,u_5|j',v')_{n_4n_5}
K\left(%
\begin{array}{cc}
  n_5,i_5,u_5 & n_3,i_3,u_3  \\
  n_4,i_4,u_4 & n_6,i_6,u_6 \\
\end{array}%
\right)G^{-1}_{n_3n_2}(i_3,u_3;i_2,u_2).\label{Fock
sigma}\end{eqnarray}
The Fock part of the fermion self-energy depends on the two-particle
Green's function $K$; therefore the SD equations and the BS equation
for $K$ have to be solved self-consistently.

Our approach to the Hubbard model allows us to obtain exact
equations of the Green's functions by using the field-theoretical
technique. We now wish to return to our statement that the Green's
functions are the thermodynamic average of the
$\widehat{T}_u$-ordered products of field operators. The standard
procedure for calculating the Green's functions, is to apply the Wick's
theorem. This  enables us to evaluate the $\widehat{T}_u$-ordered
products of field operators as a perturbation expansion involving
only wholly contracted field operators. These expansions can be
summed formally to yield different equations of Green's functions.
The main disadvantage of this procedure is that the validity of the
equations must be verified diagram by diagram. For this reason we
shall  use the method of Legendre transforms of the generating
functional for connected Green's functions. By applying the same
steps as  in  Ref. [\onlinecite{ZKexc}], we obtain the BS equation of the
two-particle Green's function, the Dyson equation of the boson
Green's function, and the vertex equation:
\begin{equation}
 K^{-1}\left(%
\begin{array}{cc}
  n_2,i_2,u_2 & n_3,i_3,u_3  \\
  n_1,i_1,u_1 & n_4,i_4,u_4 \\
\end{array}%
\right)= K^{(0)-1}\left(%
\begin{array}{cc}
  n_2,i_2,u_2 & n_3,i_3,u_3  \\
  n_1,i_1,u_1 & n_4,i_4,u_4 \\
\end{array}%
\right)-I\left(%
\begin{array}{cc}
  n_2,i_2,u_2 & n_3,i_3,u_3  \\
  n_1,i_1,u_1 & n_4,i_4,u_4 \\
\end{array}%
\right),\label{BSK}
\end{equation}
\begin{equation}
D_{\alpha \beta}(z,z')=D^{(0)}_{\alpha \beta}(z,z')+D^{(0)}_{\alpha
\gamma}(z,z'')\Pi_{\gamma\delta}(z'',z''')D^{(0)}_{\delta
\beta}(z,z'),\label{BosonDyson}\end{equation}
\begin{equation}\begin{split}
&\widehat{\Gamma}_{\alpha}(i_2,u_2;i_1,u_1\mid
z)_{n_2n_1}=\widehat{\Gamma}^{(0)}_{\alpha}(i_2,u_2;i_1,u_1\mid
z)_{n_2n_1}+I\left(%
\begin{array}{cc}
 n_2,i_2,u_2 & n_3,i_3,u_3  \\
  n_1,i_1,u_1 & n_4,i_4,u_4 \\
\end{array}%
\right)\times\nonumber\\&
K^{(0)}\left(%
\begin{array}{cc}
 n_3,i_3,u_3 & n_6,i_6,u_6  \\
  n_4,i_4,u_4 & n_5,i_5,u_5 \\
\end{array}%
\right)\widehat{\Gamma}_{\alpha}(i_6,u_6;i_5,u_5\mid z)_{n_6n_5}.
\label{Edward}\end{split}\end{equation}
Here, $$K^{(0)}\left(%
\begin{array}{cc}
 n_2,i_2,u_2 & n_3,i_3,u_3  \\
  n_1,i_1,u_1 & n_4,i_4,u_4 \\
\end{array}%
\right)=G_{n_2n_3}(i_2,u_3;i_2,u_2)G_{n_4n_1}(i_4,u_4;i_1,u_1)$$  is
the two-particle free propagator constructed from a pair of fully
dressed generalized single-particle Green's functions. The kernel
$I=\delta\Sigma/\delta G$ of the BS equation can be expressed as a
functional derivative  of the fermion self-energy
$\widehat{\Sigma}$. Since
$\widehat{\Sigma}=\widehat{\Sigma}^H+\widehat{\Sigma}^F$, the BS
kernel $I=I_{exc}+I_d$ is a sum of
 functional derivatives of the Hartree $\Sigma^H$ and
Fock $\Sigma^F$ contributions to the self-energy:
\begin{equation}
I_{exc}\left(%
\begin{array}{cc}
 n_2,i_2,u_2 & n_3,i_3,u_3  \\
  n_1,i_1,u_1 & n_4,i_4,u_4 \\
\end{array}%
\right)=\frac{\delta\Sigma^H(i_2,u_2;i_1,u_1)_{n_2n_1}}{\delta
G_{n_3n_4}(i_3,u_3;i_4,u_4)},\quad
I_d\left(%
\begin{array}{cc}
 n_2,i_2,u_2 & n_3,i_3,u_3  \\
  n_1,i_1,u_1 & n_4,i_4,u_4 \\
\end{array}%
\right)=\frac{\delta\Sigma^F(i_2,u_2;i_1,u_1)_{n_2n_1}}{\delta
G_{n_3n_4}(i_3,u_3;i_4,u_4)}.\label{Kernel}\end{equation} The
general response function $\Pi$ in the Dyson equation
(\ref{BosonDyson}) is defined as
\begin{equation}\Pi_{\alpha \beta}(z,z')=
\widehat{\Gamma}^{(0)}_{\alpha}(i_1,u_1;i_2,u_2 \mid
z)_{n_1n_2}K\left(%
\begin{array}{cc}
  n_2,i_2,u_2 & n_3,i_3,u_3  \\
  n_1,i_1,u_1 & n_4,i_4,u_4 \\
\end{array}%
\right)\widehat{\Gamma}^{(0)}_{\beta}(i_3,u_3,i_4,u_4\mid
z')_{n_3n_4} .\label{Pi}
\end{equation}
 The  functions $D$, $K$  and $\widehat{\Gamma}$
 are related by the  identity:
\begin{equation}\begin{split}&
K^{(0)}\left(%
\begin{array}{cc}
 n_2,i_2,u_2 & n_3,i_3,u_3  \\
  n_1,i_1,u_1 & n_4,i_4,u_4 \\
\end{array}%
\right)\widehat{\Gamma}_{\beta}(i_4,u_4;i_3,u_3\mid
z')_{n_4n_3}D_{\beta \alpha}(z',z)\\&= K\left(%
\begin{array}{cc}
  n_2,i_2,u_2 & n_3,i_3,u_3  \\
  n_1,i_1,u_1 & n_4,i_4,u_4 \\
\end{array}%
\right)\widehat{\Gamma}^{(0)}_{\beta}(i_4,u_4;i_3,u_3\mid
z')_{n_4n_3}D^{(0)}_{\beta \alpha}(z',z), \label{GammaK}
\end{split}\end{equation}

 By introducing the boson proper self-energy $P^{-1}_{\alpha
\beta}(z,z')=
\Pi^{-1}_{\alpha\beta}(z,z')+D^{(0)}_{\alpha\beta}(z,z')$, one can
rewrite the Dyson equation (\ref{BosonDyson})
 for the boson Green's function as:
\begin{equation}D^{-1}_{\alpha \beta}(z,z') =D^{(0)-1}_{\alpha
\beta}(z,z')-P_{\alpha \beta}(z,z').\label{DE}\end{equation} The
proper self-energy and the vertex function $\widehat{\Gamma}$ are
related by the following equation:
 \begin{equation}\begin{split}&
P_{\alpha
\beta}(z,z')=\frac{1}{2}Tr\left[\widehat{\Gamma}_\alpha^{(0)}(y_1,x_2|z)\widehat{G}(x_2,y_3)
\widehat{\Gamma}_\beta(y_3,x_4|z')\widehat{G}(x_4,y_1)\right]\nonumber\\&=\frac{1}{2}
\widehat{\Gamma}^{(0)}_{\alpha}(i_1,u_1;i_2,u_2\mid
z)_{n_1n_2}G_{n_2n_3}(i_2,u_2;i_3,u_3)
\widehat{\Gamma}_{\beta}(i_3,u_3;i_4,u_4\mid
z')_{n_3n_4}G_{n_4n_1}(i_4,u_4;i_1,u_1).
\label{PG}\end{split}\end{equation}It is also possible to express
the proper self-energy in terms of the two-particle Green's function
$\widetilde{K}$ which satisfies the BS equation
$\widetilde{K}^{-1}=K^{(0)-1}-I_d$, but its kernel
$I_d=\delta\Sigma^F/\delta G$ includes only diagrams that represent
the direct interactions:
  \begin{equation}\begin{split}&
P_{\alpha
\beta}(z,z')=\widehat{\Gamma}^{(0)}_{\alpha}(i_1,u_1;i_2,u_2\mid
z)_{n_1n_2}\widetilde{K}\left(%
\begin{array}{cc}
  n_2,i_2,u_2 & n_3,i_3,u_3  \\
  n_1,i_1,u_1 & n_4,i_4,u_4 \\
\end{array}%
\right) \widehat{\Gamma}^{(0)}_{\beta}(i_3,u_3;i_4,u_4\mid
z')_{n_3n_4}\nonumber\\&
=\widehat{\Gamma}^{(0)}_{n_1n_2}(\alpha)\widetilde{K}\left(%
\begin{array}{cc}
  n_2,\textbf{r}_j,v & n_3,\textbf{r}_{j'},v'  \\
  n_1,\textbf{r}_j,v & n_4,\textbf{r}_{j'},v' \\
\end{array}\right)\widehat{\Gamma}^{(0)}_{n_3n_4}(\beta). \label{P}\end{split}\end{equation}
One can obtain the spectrum of the collective excitations as poles of the boson Green's function by solving the Dyson equation (\ref{DE}), but  first we have to deal with the BS equation for the function $\widetilde{K}$. In other words, this method for obtaining the collective modes requires two steps. For this reason, it is easy to  obtain the collective modes by locating the poles of the two-particle Green's function $K$ using the solutions of the corresponding BS equation.

  As we have already mentioned, the BS equation and
the SD equations  have to be solved self-consistently. In what follows, we use an approximation which allows us to decouple the
above-mentioned equations and to obtain a linearized integral
equation for the Fock term. To apply  this approximation, we first
use Eq. (\ref{GammaK}) to rewrite the Fock term as
\begin{equation}\Sigma^F(i_1,u_1;i_2,u_2)_{n_1n_2}=-
\widehat{\Gamma}_{\alpha}^{(0)}(i_1,u_1;i_3,u_3|j,v)_{n_1n_3}
D_{\alpha\beta}(j,v;j',v')G_{n_3n_4}(i_3,u_3;i_4,u_4)\widehat{\Gamma}_{\beta}(i_4,u_4;i_2,u_2|j',v')_{n_4n_2}
,\label{MassFock}\end{equation} and after that, we replace $D$ and
$\widehat{\Gamma}$ in (\ref{MassFock}) by the free boson propagator
$D^{(0)}$  and by the bare vertex $\widehat{\Gamma}^{(0)}$,
respectively.  In this approximation the Fock term assumes the form:
 \begin{equation}\begin{split}
&\Sigma_0^F(i_1,u_1;i_2,u_2)_{n_1n_2}=-
\widehat{\Gamma}_{\alpha}^{(0)}(i_1,u_1;i_3,u_3|j,v)_{n_1n_3}
D^{(0)}_{\alpha\beta}(j,v;j',v')\widehat{\Gamma}_{\beta}^{(0)}(i_4,u_4;i_2,u_2|j',v'))_{n_4n_2}
G_{n_3n_4}(i_3,u_3;i_4,u_4)=\\&
-U\delta_{i_1,i_2}\delta(u_1-u_2)\left(%
\begin{array}{cccc}
  0 &G_{12}(1;2)&0&-G_{14}(1;2)  \\
 G_{21}(1;2)& 0 &-G_{23}(1;2)&0  \\
 0 & -G_{32}(1;2)&0&G_{34}(1;2)  \\
 -G_{41}(1;2) & 0&G_{43}(1;2)&0  \\
\end{array}%
\right).\label{SF1}
\end{split}\end{equation}
The total self-energy is $\widehat{\Sigma}(i_1,u_1;i_2,u_2)=\widehat{\Sigma}^H(i_1,u_1;i_2,u_2)+\widehat{\Sigma}^F(i_1,u_1;i_2,u_2)$, where
\begin{equation}
\widehat{\Sigma}^H(i_1,u_1;i_2,u_2)=\frac{U}{2}\delta_{i_1,i_2}\delta(u_1-u_2)\left(%
\begin{array}{cccc}
  G_{22}-G_{44} &0&0&0 \\
 0& G_{11}-G_{33} &0&0  \\
 0 & 0&G_{44}-G_{22}&0  \\
 0& 0&0&G_{33}-G_{11}  \\
\end{array}%
\right)\label{HFG},\end{equation}
\end{widetext}
where $G_{ij}\equiv G_{ij}(1;2)=G_{ij}(i_1,u_1;i_2,u_2)$.

The
contributions to $\Sigma(i_1,u_1;i_2,u_2)$ due to the elements on
the major diagonal of the above matrices can be included into the
chemical potential. To obtain an analytical
expression for the generalized single-particle Green's function, we  assume
two more approximations. First, since the experimentally relevant magnetic fields are not strong enough to cause
spin flips, we shall assume
$G_{12}=G_{21}=G_{34}=G_{43}=0$. Second, we neglect the
frequency dependence of the Fourier transform of the Fock part of
the fermion self-energy. Thus, the Dyson equation for the
generalized single-particle Green's function becomes:
\begin{equation}\widehat{G}^{-1}(1;2)=\left(%
\begin{array}{cccc}
  G^{(0)-1}_{11} &G^{(0)-1}_{12}&0&-\Delta_{i_1,i_2}  \\
 G^{(0)-1}_{21}& G^{(0)-1}_{22} &\Delta_{i_1,i_2}&0  \\
 0 & \Delta_{i_1,i_2} &G^{(0)-1}_{33}&G^{(0)-1}_{31}  \\
 -\Delta_{i_1,i_2} & 0&G^{(0)-1}_{41}&G^{(0)-1}_{44}  \\
\end{array}%
\right),\label{GFZ}\end{equation}
where $G^{(0)-1}_{ij}=G^{(0)-1}_{ij}(1;2)$. In the
population-balanced case $\Delta_{i_1,i_2}\equiv \Delta_{0} \delta(\textbf{r}_{i_1}-\textbf{r}_{i_2}) $, where $\Delta_{0}$ is  a constant in space. Physically, it describes a superfluid state of  Cooper
pairs with zero momentum. Superfluid states of Cooper pairs with
nonzero momentum  occur in population-imbalanced case between a
fermion with momentum $\textbf{k} + \textbf{q}$ and spin $\uparrow$
and a fermion with momentum $-\textbf{k} + \textbf{q}$, and spin
$\downarrow$ . As a result, the pair momentum is $2\textbf{q}$. A
finite pairing momentum implies a position-dependent phase of the
order parameter, which in the  Fulde-Ferrell  (FF)
case  varies as a single plane wave  $\Delta_{i_1,i_2}\equiv \Delta\textbf{q} e^{\imath2\textbf{q.r}_{i_1}}\delta(\textbf{r}_{i_1}-\textbf{r}_{i_2}) $.\cite{FF}
\subsection{Mean field approximation}
The poles of the mean field single-particle Green's function of the FF superfluidity in the present of SOC and the Zeeman field are defined by very long expressions. The numerical solution of the mean field  set of two number equations, the gap equation and the equation for the FF vector $\textbf{q}$ is an ambitious task which will be left as a subject of our future research.

In what follows, we  consider only pairing between
atoms with equal and opposite momenta, i.e. the BCS superfluid in the balanced system, and the Sarma\cite{Sarma} superfluid in the imbalanced case. In the imbalanced  case,  the Fourier transform of  the zero-temperature single-particle Green's function (\ref{GFZ}) in the mean field approximation is given by\cite{GSOC6}
\begin{widetext}
     \begin{equation}
 \widehat{G}_{MF}^{-1}(\textbf{k},\imath\omega_m)=\left(%
\begin{array}{cccc}
  \imath\omega_m-\xi_\uparrow(\textbf{k})-h_z&-2\lambda\left(\sin k_x+\imath\sin k_y\right)&0&\Delta_0  \\
  -2\lambda\left(\sin k_x-\imath\sin k_y\right)&\imath\omega_m-\xi_\downarrow(\textbf{k})+h_z &-\Delta_0&0  \\
 0 &-\Delta_0 &\imath\omega_m+\xi_\uparrow(\textbf{k})+h_z&-2\lambda\left(\sin k_x-\imath\sin k_y\right)\\
 \Delta_0 &0 & -2\lambda\left(\sin k_x+\imath\sin k_y\right)&\imath\omega_m+\xi_\downarrow(\textbf{k})-h_z
\end{array}%
\right).
 \label{GFMF}\end{equation}
Here, the chemical potentials $\mu_{\uparrow,\downarrow}$ and the gap $\Delta_0$ are defined by the solutions of the mean-field number and gap equations:
\begin{equation}\begin{split}
&n=f_\uparrow+f_\downarrow=1-\sum_{\textbf{k}}\left[\frac{1}{2}-f(\Omega(\textbf{k}))\right]\frac{\chi(\textbf{k})}{\Omega(\textbf{k})}
\left(1+\frac{S(\textbf{k})+\eta^2(\textbf{k})}{\sqrt{\left(S(\textbf{k})
+\eta^2(\textbf{k})\right)\left(\chi^2(\textbf{k})+\Delta^2_0\right)-S(\textbf{k})\Delta_0^2}}\right)\\&
-\sum_{\textbf{k}}\left[\frac{1}{2}-f(\omega(\textbf{k}))\right]\frac{\chi(\textbf{k})}{\omega(\textbf{k})}
\left(1-\frac{S(\textbf{k})+\eta^2(\textbf{k})}{\sqrt{\left(S(\textbf{k})
+\eta^2(\textbf{k})\right)\left(\chi^2(\textbf{k})+\Delta^2_0\right)-S(\textbf{k})\Delta_0^2}}\right)
\label{N1}\end{split}\end{equation}
\begin{equation}\begin{split}
&nP=f_\uparrow-f_\downarrow=-\sum_{\textbf{k}}\left[\frac{1}{2}-f(\Omega(\textbf{k}))\right]\frac{\eta(\textbf{k})}{\Omega(\textbf{k})}
\left(1+\frac{\chi^2(\textbf{k})+\Delta_0^2}{\sqrt{\left(S(\textbf{k})
+\eta^2(\textbf{k})\right)\left(\chi^2(\textbf{k})+\Delta^2_0\right)-S(\textbf{k})\Delta_0^2}}\right)\\&
-\sum_{\textbf{k}}\left[\frac{1}{2}-f(\omega(\textbf{k}))\right]\frac{\eta(\textbf{k})}{\omega(\textbf{k})}
\left(1-\frac{\chi^2(\textbf{k})+\Delta_0^2}{\sqrt{\left(S(\textbf{k})
+\eta^2(\textbf{k})\right)\left(\chi^2(\textbf{k})+\Delta^2_0\right)-S(\textbf{k})\Delta_0^2}}\right),
\label{N2}\end{split}\end{equation}
\begin{equation}\begin{split}
&\frac{1}{U}=\sum_{\textbf{k}}\left[\frac{1}{2}-f(\Omega(\textbf{k}))\right]\frac{1}{2\Omega(\textbf{k})}
\left(1+\frac{\eta^2(\textbf{k})}{\sqrt{\left(S(\textbf{k})
+\eta^2(\textbf{k})\right)\left(\chi^2(\textbf{k})+\Delta^2_0\right)-S(\textbf{k})\Delta_0^2}}\right)\\&
+\sum_{\textbf{k}}\left[\frac{1}{2}-f(\omega(\textbf{k}))\right]\frac{1}{2\omega(\textbf{k})}
\left(1-\frac{\eta^2(\textbf{k})}{\sqrt{\left(S(\textbf{k})
+\eta^2(\textbf{k})\right)\left(\chi^2(\textbf{k})+\Delta^2_0\right)-S(\textbf{k})\Delta_0^2}}\right).
\label{gap}\end{split}\end{equation}
Here, $P=(f_\uparrow-f_\downarrow)/(f_\uparrow+f_\downarrow)$ is the polarization,  $f(\omega)$ is the Fermi-Dirac distribution function, and  the following notations have been introduced:
$$\chi(\textbf{k})=\frac{1}{2}\left(\xi_{\uparrow}(\textbf{k})+\xi_{\downarrow}(\textbf{k})\right),\quad \eta(\textbf{k})=\frac{1}{2}\left(\xi_{\uparrow}(\textbf{k})-\xi_{\downarrow}(\textbf{k})\right)+h_z,\quad S(\textbf{k})=4\lambda^2\left(\sin^2k_x+\sin^2k_y\right).$$
The single-particle spectrum is
$$\Omega(\textbf{k})=\pm\sqrt{S(\textbf{k})+\frac{1}{2}\left(\omega^2_+(\textbf{k})+\omega^2_-(\textbf{k})\right)+
2\sqrt{\left(S(\textbf{k})
+\eta^2(\textbf{k})\right)\left(\chi^2(\textbf{k})+\Delta^2_0\right)-S(\textbf{k})\Delta_0^2}},$$ $$\omega(\textbf{k})=\pm\sqrt{S(\textbf{k})+\frac{1}{2}\left(\omega^2_+(\textbf{k})+\omega^2_-(\textbf{k})\right)-
2\sqrt{\left(S(\textbf{k})
+\eta^2(\textbf{k})\right)\left(\chi^2(\textbf{k})+\Delta^2_0\right)-S(\textbf{k})\Delta_0^2}},$$
where $\omega_{\pm}(\textbf{k})=\sqrt{\chi^2(\textbf{k})+\Delta^2_0}\pm \eta(\textbf{k})$.

 In the mean-field approximation, the components $G^{MF}_{n_1n_2}$ ($n_1,n_2=1,2,3,4$) of the zero-temperature  single-particle Green's function $ \widehat{G}_{MF}(\textbf{k},\omega)$ are given by
\begin{equation}
G^{MF}_{n_1n_2}(\textbf{k},\omega)=\frac{A_{n_1n_2}(\textbf{k})}{\omega-\Omega(\textbf{k})+\imath 0^+}+\frac{B_{n_1n_2}(\textbf{k})}{\omega+\Omega(\textbf{k})-\imath 0^+}+\frac{C_{n_1n_2}(\textbf{k})}{\omega-\omega(\textbf{k})+\imath 0^+}+\frac{D_{n_1n_2}(\textbf{k})}{\omega+\omega(\textbf{k})-\imath 0^+},\label{SPGFMF}
\end{equation}
where the corresponding expressions for $A_{n_1n_2}(\textbf{k}),B_{n_1n_2}(\textbf{k}),C_{n_1n_2}(\textbf{k})$ and $D_{n_1n_2}(\textbf{k})$ can be obtained by inverting the matrix (\ref{GFMF}).  As an example, we have provided in the Appendix $A_{11}(\textbf{k}),B_{11}(\textbf{k}),C_{11}(\textbf{k})$ and $D_{11}(\textbf{k})$.
\subsection{The Bethe-Salpeter equation for the collective excitations  in the generalized random phase approximation}
The spectrum of the collective modes will be obtained by solving the BS equation  in the GRPA. As we
have already mentioned, the kernel of the BS equation is a sum of
the direct $I_d=\delta \Sigma^F /\delta G$ and exchange
$I_{exc}=\delta \Sigma^H /\delta G$ interactions, written as
derivatives of the Fock (\ref{SF1}) and the Hartree (\ref{HFG}) parts of the self-energy. Thus, in the GRPA  the corresponding equation for the
BS amplitude $\Psi^{\textbf{Q}}_{n_2,n_1}$
 is given by

\begin{equation}\Psi^{\textbf{Q}}_{n_2n_1}=K^{(0)}\left(%
\begin{array}{cc}
  n_1 & n_3  \\
  n_2 & n_4 \\
\end{array}%
|\omega(\textbf{Q})\right)
\left[I_d\left(%
\begin{array}{cc}
  n_3 & n_5  \\
  n_4 & n_6 \\
\end{array}%
\right)+I_{exc}\left(%
\begin{array}{cc}
  n_3 & n_5  \\
  n_4 & n_6 \\
\end{array}%
\right)\right]\Psi^{\textbf{Q}}_{n_6,n_5},\label{BSEdZ1}
\end{equation}
where $I_d\left(%
\begin{array}{cc}
  n_1 & n_3  \\
  n_2 & n_4 \\
\end{array}%
\right)=-\Gamma^{(0)}_\alpha(n_1,n_3)D^{(0)}_{\alpha\beta}
\Gamma^{(0)}_\beta(n_4,n_2)$ and $I_{exc}\left(%
\begin{array}{cc}
  n_1 & n_3  \\
  n_2 & n_4 \\
\end{array}%
\right)=
\frac{1}{2}\Gamma^{(0)}_\alpha(n_1,n_2)D^{(0)}_{\alpha\beta}
\Gamma^{(0)}_\beta(n_4,n_3)$ are the direct and the exchange interactions, correspondingly. The two-particle propagator $K^{(0)}$ in the GRPA is defined  as follows:
\begin{equation}\begin{split}&K^{(0)}\left(%
\begin{array}{cc}
  n_1 & n_3  \\
  n_2 & n_4 \\
\end{array}%
|\omega(\textbf{Q})\right)\equiv K_{n_1n_3n_4n_2}=\int
\frac{d\Omega}{2\pi}
 \int\frac{d^d\textbf{k}}{(2\pi)^d}G^{MF}_{n_1n_3}
 \left(\textbf{k}+\textbf{Q},\Omega+\omega(\textbf{Q})\right)G^{MF}_{n_4n_2}(\textbf{k},\Omega)\\&
 =\frac{A_{n_1n_3}(\textbf{k}+\textbf{Q})B_{n_4n_2}(\textbf{k})}
 {\omega-\left(\Omega(\textbf{k}+\textbf{Q})+\Omega(\textbf{k})\right)}-\frac{B_{n_1n_3}(\textbf{k}+\textbf{Q})A_{n_4n_2}(\textbf{k})}
 {\omega+\left(\Omega(\textbf{k}+\textbf{Q})+\Omega(\textbf{k})\right)}+\frac{C_{n_1n_3}(\textbf{k}+\textbf{Q})D_{n_4n_2}(\textbf{k})}
 {\omega-\left(\omega(\textbf{k}+\textbf{Q})+\omega(\textbf{k})\right)}
 -\frac{D_{n_1n_3}(\textbf{k}+\textbf{Q})C_{n_4n_2}(\textbf{k})}
 {\omega+\left(\omega(\textbf{k}+\textbf{Q})+\omega(\textbf{k})\right)}\\&+\frac{A_{n_1n_3}(\textbf{k}+\textbf{Q})D_{n_4n_2}(\textbf{k})}
 {\omega-\left(\Omega(\textbf{k}+\textbf{Q})+\omega(\textbf{k})\right)}-\frac{B_{n_1n_3}(\textbf{k}+\textbf{Q})C_{n_4n_2}(\textbf{k})}
 {\omega+\left(\Omega(\textbf{k}+\textbf{Q})+\omega(\textbf{k})\right)}+
 \frac{C_{n_1n_3}(\textbf{k}+\textbf{Q})B_{n_4n_2}(\textbf{k})}
 {\omega-\left(\omega(\textbf{k}+\textbf{Q})+\Omega(\textbf{k})\right)}
 -\frac{D_{n_1n_3}(\textbf{k}+\textbf{Q})A_{n_4n_2}(\textbf{k})}
 {\omega+\left(\omega(\textbf{k}+\textbf{Q})+\Omega(\textbf{k})\right)},\label{BSGRPA}
 \end{split}\end{equation}
The BS equation (\ref{BSEdZ1}) can be  written in the matrix form as
$\left(\widehat{I}+U\widehat{Z}\right)\widehat{\Psi}=0$, where
 $\widehat{I}$ is the unit matrix,  the matrix $\widehat{Z}$ is a $16 \times 16$ matrix, and
 the transposed matrix of
$\widehat{\Psi}$  is given by:
$$\widehat{\Psi}^T=
(\Psi^{\textbf{Q}}_{1,1} \quad
  \Psi^{\textbf{Q}}_{1,2} \quad
  \Psi^{\textbf{Q}}_{1,3} \quad
  \Psi^{\textbf{Q}}_{1,4} \quad
  \Psi^{\textbf{Q}}_{2,1} \quad
  \Psi^{\textbf{Q}}_{2,2} \quad
  \Psi^{\textbf{Q}}_{2,3}\quad
  \Psi^{\textbf{Q}}_{2,4} \quad
  \Psi^{\textbf{Q}}_{3,1} \quad
  \Psi^{\textbf{Q}}_{3,2} \quad
  \Psi^{\textbf{Q}}_{3,3} \quad
  \Psi^{\textbf{Q}}_{3,4} \quad
  \Psi^{\textbf{Q}}_{4,1} \quad
  \Psi^{\textbf{Q}}_{4,2} \quad
  \Psi^{\textbf{Q}}_{4,3} \quad
  \Psi^{\textbf{Q}}_{4,4}
).
$$
The collective-mode dispersion, $\omega(\textbf{Q})$, follows from the condition that the  secular determinant $det|\widehat{I}+U\widehat{Z}|$ is equal to zero. By standard algebraic manipulations, the $16 \times 16$  determinant, that follows from the BS equation, can be reduced to the following $12 \times 12$ secular determinant:
\begin{eqnarray}&
Z_{12\times 12}=\left|%
\begin{array}{ccccccccc}
1+U/2\left(K_{1221}-K_{1441}\right)&-UK_{1211}&UK_{1411}&-UK_{1121}&U/2\left(K_{1111}-K_{1331}\right)&UK_{1321}\\
 U/2\left(K_{2221}-K_{2441}\right)&1-UK_{2211}&UK_{2411}&-UK_{2121}&U/2\left(K_{2111}-K_{2331}\right)&UK_{2321}\\
U/2\left(K_{4221}-K_{4441}\right)&-UK_{4211}&1+UK_{4411}&-UK_{4121}&U/2\left(K_{4111}-K_{4331}\right)&UK_{4321}\\
U/2\left(K_{1222}-K_{1442}\right)&-UK_{1212}&UK_{1412}&1-UK_{1122}&U/2\left(K_{1112}-K_{1332}\right)&UK_{1322}\\
U/2\left(K_{1222}-K_{1442}\right)&-UK_{2212}&UK_{2412}&-UK_{2122}&1+U/2\left(K_{2112}-K_{2332}\right)&UK_{2322}\\
U/2\left(K_{3222}-K_{3442}\right)&-UK_{3212}&UK_{3412}&-UK_{3122}&U/2\left(K_{3112}-K_{3332}\right)&1+UK_{3322}\\
 U/2\left(K_{2223}-K_{2443}\right)&-UK_{2213}&UK_{2413}&-UK_{2123}&U/2\left(K_{2113}-K_{2333}\right)&UK_{2323}\\
 U/2\left(K_{3223}-K_{3443}\right)&UK_{3213}&UK_{3413}&-UK_{3123}&U/2\left(K_{3113}-K_{3333}\right)&UK_{3323}\\
U/2\left(K_{4223}-K_{4443}\right)&-UK_{4213}&UK_{4413}&-UK_{4123}&U/2\left(K_{4113}-K_{4333}\right)&UK_{4323}\\
U/2\left(K_{1224}-K_{1444}\right)&-UK_{1214}&UK_{1414}&-UK_{1124}&U/2\left(K_{1114}-K_{1334}\right)&UK_{1324}\\
U/2\left(K_{3224}-K_{3444}\right)&-UK_{3214}&UK_{3414}&-UK_{3124}&U/2\left(K_{3114}-K_{3334}\right)&UK_{3324}\\
U/2\left(K_{4224}-K_{4444}\right)&-UK_{4214}&UK_{4414}&-UK_{4124}&U/2\left(K_{4114}-K_{4334}\right)&UK_{4324}\\
 \\
 \end{array}%
\right. \nonumber\\&\left.%
\begin{array}{cccccccc}
UK_{1231}&U/2\left(K_{1441}-K_{1221}\right)&-UK_{1431}&UK_{1141}&-UK_{1341}&U/2\left(K_{1331}-K_{1111}\right)\\
UK_{2231}&U/2\left(K_{2441}-K_{2221}\right)&-UK_{2431}&UK_{2141}&-UK_{2341}&U/2\left(K_{2331}-K_{2111}\right)\\
UK_{4231}&U/2\left(K_{4441}-K_{4221}\right)&-UK_{4431}&UK_{4141}&-UK_{4341}&U/2\left(K_{4331}-K_{4111}\right)\\
UK_{1232}&U/2\left(K_{1442}-K_{1222}\right)&-UK_{1432}&UK_{1142}&-UK_{1342}&U/2\left(K_{1332}-K_{1112}\right)\\
UK_{2232}&U/2\left(K_{2442}-K_{2222}\right)&-UK_{2432}&UK_{2142}&-UK_{2342}&U/2\left(K_{2332}-K_{2112}\right)\\
UK_{3232}&U/2\left(K_{3442}-K_{3222}\right)&-UK_{3432}&UK_{3142}&-UK_{3342}&U/2\left(K_{3332}-K_{3112}\right)\\
1+UK_{2233}&U/2\left(K_{2443}-K_{2223}\right)&-UK_{2433}&UK_{2143}&-UK_{2343}&U/2\left(K_{2333}-K_{2113}\right)\\
UK_{3233}&1+U/2\left(K_{3443}-K_{3223}\right)&-UK_{3433}&UK_{3143}&-UK_{3343}&U/2\left(K_{3333}-K_{3113}\right)\\
UK_{4233}&U/2\left(K_{4443}-K_{4223}\right)&1-UK_{4433}&UK_{4143}&-UK_{4343}&U/2\left(K_{4333}-K_{4113}\right)\\
UK_{1234}&U/2\left(K_{1444}-K_{1224}\right)&-UK_{1434}&1+UK_{1144}&-UK_{1344}&U/2\left(K_{1334}-K_{1114}\right)\\
UK_{3234}&U/2\left(K_{3444}-K_{3224}\right)&-UK_{3434}&UK_{3144}&1-UK_{3344}&U/2\left(K_{3334}-K_{3114}\right)\\
UK_{4234}&U/2\left(K_{4444}-K_{4224}\right)&-UK_{4434}&UK_{4144}&-UK_{4344}&1+U/2\left(K_{4334}-K_{4114}\right)\\
 \\
 \end{array}%
\right| \label{FF88}\end{eqnarray}
In the above secular determinant, there are 144 two-particle Green's functions, but only 78 of them are different: $K_{1111}$, $K_{1144}$, $K_{1122}$, $K_{2211}$, $K_{2222}$, $K_{2233}$, $K_{3322}$, $K_{3333}$, $K_{3344}$, $K_{4433}$, $K_{4411}$, $K_{4444}$, and 66 different $K_{n_1,n_2,n_3,n_4}$ ($K_{n_1,n_2,n_3,n_4}=K_{n_2,n_1,n_4,n_3}$). It is worth mentioning that within the Gaussian approximation\cite{GSOC6,GSOC7} the secular determinant includes only 8 of them, $K_{2233}$, $K_{3322}$, $K_{1144}$, $K_{4411}$, $K_{1234}$, $K_{3421}$, $K_{2413}$, and $K_{1324}$:
\begin{equation}
Z_{2\times 2}=\left|%
\begin{array}{cc}
  1+\frac{U}{2}\left(K_{2233}+K_{1144}-K_{1234}-K_{2143}\right) & \frac{U}{2}\left(K_{1414}+K_{2323}-K_{2413}-K_{1324}\right)  \\
  \frac{U}{2}\left(K_{1414}+K_{2323}-K_{4231}-K_{3142}\right) &  1+\frac{U}{2}\left(K_{3322}+K_{4411}-K_{3412}-K_{4321}\right)  \\
\end{array}%
\right|\label{GA}\end{equation}
\end{widetext}
In next Section, we shall see that the speeds  of sound in the case of Sarma superfluid, calculated within the Gaussian approximation and from the BS equation in the GRPA, are similar, but the dispersions around the roton minimum, provided by  the two methods are remarkably different.
\section{Numerical results}
\subsection{Collective modes of an imbalanced  2D SOC Sarma superfluid in an optical lattice}
We use the secular determinants (\ref{FF88}) and (\ref{GA}) to calculate the collective-mode dispersion and the speed of sound of an imbalanced Fermi gas in a 2D
square optical lattice (lattice constant $a$) with  a Rashba SOC.   We consider  a weak coupling limit $U=2.64J$, where the BS equation in the GRPA provides a good approximation for the collective-mode energies. The strength of the SOC is $\lambda=0.1J$, and the system parameters are chosen so that the density of the majority and minority components are $f_\uparrow=0.275$  and $f_\downarrow=0.225$, respectively (the corresponding polarization is $P=0.1$). The chemical potentials $\mu_\uparrow=2.857J$, $\mu_\downarrow=2.186J$ and the gap $\Delta=0.266J$ are obtained by numerically solving the number and the gap equations (\ref{N1})-(\ref{gap}).

\begin{figure}
\includegraphics[scale=0.65]{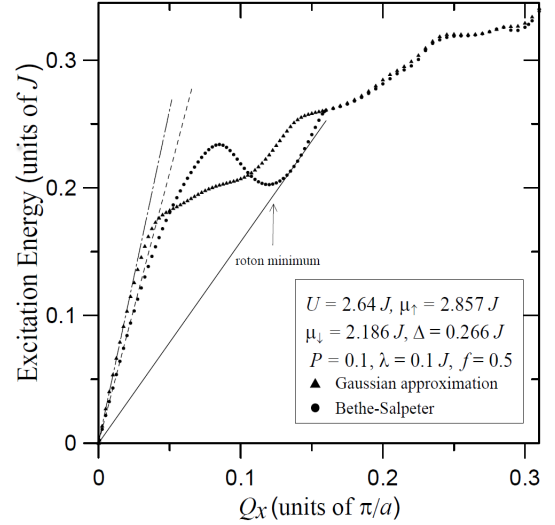}
    \caption{Collective modes dispersion $\omega(Q_x)$  of an imbalanced SOC Fermi gas in a 2D square optical lattice along the $(Q_x,0)$ direction, obtained by the Gaussian approximation (triangles) and the Bethe-Salpeter equation in the GRPA (circles).}\label{Fig3}\end{figure}

In Fig. \ref{Fig3}, we plot the collective-mode excitation energy $\omega(Q_x)$  as a function of the wave vector $\textbf{Q}=(Q_x,0)$ along the x-axis, calculated by the Gaussian approximation (triangles) and by the  BS approach in the GRPA (circles). The speed of sound,  provided by the Gaussian approximation,  is $u =$ 1.66 $Ja/\hbar$, while the speed of sound, calculated within the BS approach is  $u =$ 1.35 $Ja/\hbar$. Thus, the Gaussian approximation overestimates the speed of sound by about 23\%.  More interestingly, the dispersion curve calculated from the BS equation clearly shows the existence of a roton-like minimum, while there is no such a minimum within the Gaussian approximation. The corresponding roton gap is $\Delta_r = 0.2025 J$ and the critical flow velocity, obtained around the roton minimum from the BS equation, is $v_c =$ 0.51 $Ja/\hbar$. As in 2D case, shown in Fig. \ref{Fig2}, the two dispersion curves are remarkably different around the roton minimum; instead of the expected roton-like structure, the dispersion curve  provided by the Gaussian approximation monotonically increases in this region. At higher wave vectors, $Q_x >$ 0.16 $\pi/a$, the two dispersion curves have essentially the same behavior.

Our results suggest that the Gaussian approximation overestimates the speed of sound of the Goldstone mode, and  fails to reproduced the roton-like structure of the collective-mode dispersion which appears after the linear part of the dispersion. The question naturally arises here, whether the Gaussian approximation still can be used to estimate the speed of sound if one takes into account not only the SOC, but Zeeman fields as well.  This question will be answered in the next subsection, where we investigate the speed of sound of a balanced superfluid Fermi gas  in a two-dimensional square optical lattice with a Rashba spin-orbit coupling (in the xy plane) and an out-of-plane Zeeman field.
\begin{figure}
\includegraphics[scale=0.7]{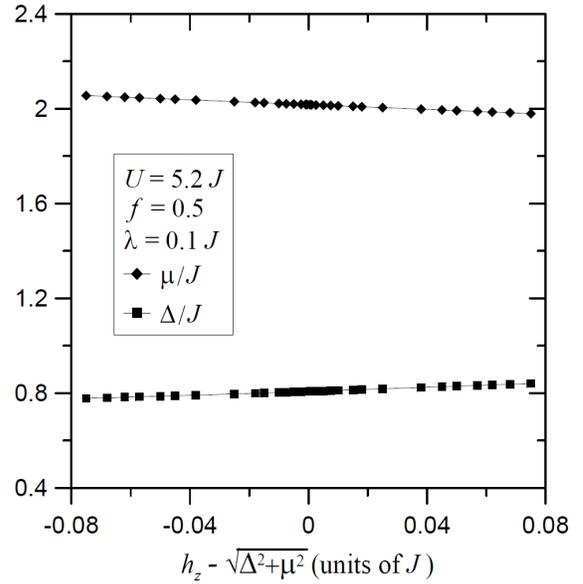}
    \caption{The chemical potential (diamonds) and the gap (squares) of a balanced superfluid Fermi gas  in a 2D square optical lattice as a function of the out-of-plane Zeeman field. The system parameters are:  filling factor  $f=0.5$, the attractive interaction $U=5.2J$, and the strength of the Rashba spin-orbit-coupling  $\lambda=0.1J$.}\label{Fig4}\end{figure}
    \begin{figure}
\includegraphics[scale=0.7]{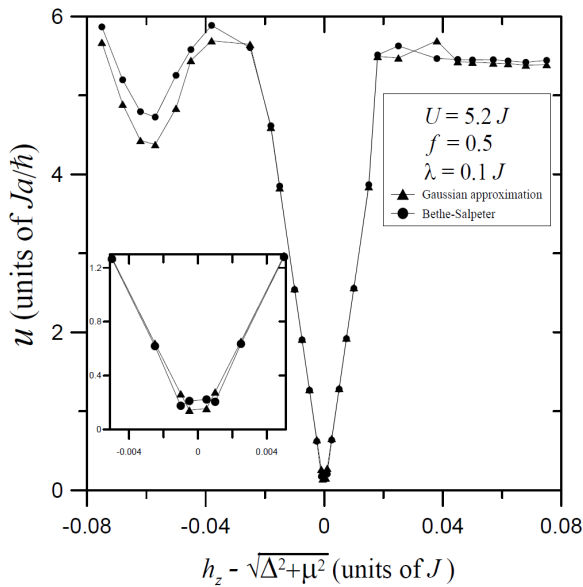}
    \caption{The speed of sound along the x direction as a function of the Zeeman field, calculated within the Gaussian approximation (circles) and by the BS approach (triangles). The values of the chemical potential and gap are shown in  Fig. \ref{Fig3}. }\label{Fig5}\end{figure}
\subsection{Speed of sound near the transition from the gapped superfluid phase to the topological phase }
  We consider a system with a filling factor $f=0.5$, and an attractive interaction $U=5.2J$. The strength of the Rashba spin-orbit-coupling is $\lambda=0.1J$, and there exists an out-of-plane Zeeman field $h_z$. In such a system a phase transition can be accessed by varying the
Zeeman field. The transition from the gapped superfluid phase to the topological phase is characterized by the
quasiparticle excitation gap that closes at $h_c=\sqrt{\mu^2+\Delta^2}$ and reopens with
increasing $h_z>h_c$.

In Fig. \ref{Fig4}, we present the chemical
potential  (diamonds) and the gap  (squares) for different Zeeman fields, obtained by solving numerically the number and the gap equations.  The critical field $h_c$ marks the phase transition
between the topologically nontrivial (negative side) and the topologically
trivial (positive side) superfluid phases. As can be seen, during this transition, the gap $\Delta$ is still finite even though the quasiparticle excitation gap is closed. This suggests that there is a quantum
phase transition  separating the parameter regimes $h_z<h_c$ and $h_z>h_c$, even though the
system in both regimes is an s-wave superfluid.

In Fig. \ref{Fig5}, we have plotted the speed of sound along x direction ($\textbf{Q}=(Q_x,0)$) as a function of the Zeeman field, calculated within the Gaussian approximation and from the BS equation in the GRPA.  As can be seen, close to the phase transition boundary the speed of sound calculated within the two approaches is essentially the same. The inset in the figure shows that the minimum of the speed of sound  is located at the phase transition boundary $h_c$. The same behavior was previously found by applying the  Gaussian approximation in the case of a 2D superfluid atomic Fermi gas with Rashba-type spin-orbit coupling and an out-of-plane Zeeman field,\cite{GSOC4,GSOC6} and in the case of a 3D  FF type of superfluid Fermi gas with Rashba spin-orbit coupling (in the xy plane) and two Zeeman fields [in-plane (hx) and out-of-plane (hz)].\cite{GSOC7} Thus, our calculations are in agreement with the  suggestion made in Ref. [\onlinecite{GSOC7}],  that by measuring the minimum of the speed of sound one can  unambiguously detect the topological phase transition boundary.

\section{Summary}
In summary, we have derived the BS equation in the GRPA for the collective excitation energy  of a Fermi gas in a 2D square lattice  with an attractive contact interaction, assuming the existence of a nonvanishing Rashba SOC and an out-of-plane Zeeman field. We have calculated the collective-mode dispersion within the Gaussian approximation, and from the BS equation assuming the existence of a Sarma superfluid state. It is found that the Gaussian approximation: (i) overestimates the speed of sound of the Goldstone mode, and (ii) fails to reproduce the roton-like structure of the collective-mode dispersion which appears in the BS approach.

We also have investigated the speed of sound of a balanced spin-orbit-coupled atomic Fermi gas near the boundary of the topological phase transition driven by an out-of-plane Zeeman field. It is shown that the minimum of the speed of sound  is located at the topological phase transition boundary, and this fact  can be used to  confirm the existence of a topological phase transition.
\begin{center}\textbf{ACKNOWLEDGEMENTS}\end{center}
This work was partially supported by the National Council for Science and Technology, CONACYT (Mexico) through the
postdoctoral grant 232171.\\\\
\begin{center}\textbf{APPENDIX}\end{center}
 The symbols $I_{a,b}$ and $J_{a,b}$ at  nonzero temperatures
are defined  as:
\begin{widetext}
\begin{equation}\begin{split}&
I_{a,b}=\frac{1}{2N}\sum_\textbf{k}a_{\textbf{k},\textbf{Q}}b_{\textbf{k},\textbf{Q}}
\left[\frac{1-f\left(\omega_-(\textbf{k})
\right)-f\left(\omega_+(\textbf{k}+\textbf{Q})
\right)}{\omega+\Omega(\textbf{k},\textbf{Q})-\varepsilon(\textbf{k},\textbf{Q})]}
-\frac{1-f\left(\omega_+(\textbf{k})
\right)-f\left(\omega_-(\textbf{k}+\textbf{Q})
\right)}{\omega+\Omega(\textbf{k},\textbf{Q})+\varepsilon(\textbf{k},\textbf{Q})]}\right]
,\\&
J_{a,b}=\frac{1}{2N}\sum_\textbf{k}a_{\textbf{k},\textbf{Q}}b_{\textbf{k},\textbf{Q}}
\left[\frac{1-f\left(\omega_-(\textbf{k})
\right)-f\left(\omega_+(\textbf{k}+\textbf{Q})
\right)}{\omega+\Omega(\textbf{k},\textbf{Q})-\varepsilon(\textbf{k},\textbf{Q})]}
+\frac{1-f\left(\omega_+(\textbf{k})
\right)-f\left(\omega_-(\textbf{k}+\textbf{Q})
\right)}{\omega+\Omega(\textbf{k},\textbf{Q})+\varepsilon(\textbf{k},\textbf{Q})]}\right]
.\nonumber\end{split}\end{equation}
Here,
$\varepsilon(\textbf{k},\textbf{Q})=
 E(\textbf{k}+\textbf{Q})+ E(\textbf{k})$,
   $\Omega(\textbf{k},\textbf{Q})=\eta(\textbf{k})-\eta
 (\textbf{k}+\textbf{Q})$, $E(\textbf{k})=\sqrt{\chi^2(\textbf{k})+\Delta^2_0}$, $\omega_{\pm}(\textbf{k})=E(\textbf{k})\pm\eta(\textbf{k})$,
 and $a$ and $b$ are one of the following form factors:
\begin{equation}\begin{split}&\gamma_{\textbf{k},\textbf{Q}}=
u_{\textbf{k}}u_{\textbf{k}+\textbf{Q}}
+v_{\textbf{k}}v_{\textbf{k}+\textbf{Q}},\quad
l_{\textbf{k},\textbf{Q}}=u_{\textbf{k}}
u_{\textbf{k}+\textbf{Q}}
-v_{\textbf{k}}v_{\textbf{k}+\textbf{Q}},
\widetilde{\gamma}_{\textbf{k},\textbf{Q}}=
u_{\textbf{k}}v_{\textbf{k}+\textbf{Q}}
-u_{\textbf{k}+\textbf{Q}}v_{\textbf{k}},\quad
 m_{\textbf{k},\textbf{Q}}=
u_{\textbf{k}}v_{\textbf{k}+\textbf{Q}}+
u_{\textbf{k}+\textbf{Q}}v_{\textbf{k}},\\&
u_{\textbf{k}}=\sqrt{\frac{1}{2}\left(1+\frac{\chi(\textbf{k})}{E(\textbf{k})}\right)},
\quad v_{\textbf{k}}=\sqrt{\frac{1}{2}\left(1-\frac{\chi(\textbf{k})}{E(\textbf{k})}\right)}
\nonumber\end{split}\end{equation}

The functions $A_{n_1n_2}(\textbf{k}),B_{n_1n_2}(\textbf{k}),C_{n_1n_2}(\textbf{k})$ and $D_{n_1n_2}(\textbf{k})$ are easily obtained by inverting the matrix (\ref{GFMF}). Here, we provide the expressions for  $A_{11}(\textbf{k}),B_{11}(\textbf{k}),C_{11}(\textbf{k})$ and $D_{11}(\textbf{k})$ (the $\textbf{k}$-dependence of   $\xi_{\uparrow,\downarrow} (\textbf{k}),S (\textbf{k}), \Omega(\textbf{k})$, and $\omega (\textbf{k})$ is not explicitly shown):
$$A_{11}(\textbf{k})=$$$$\frac{\left[-h_z+\xi_\downarrow(\textbf{k})\right]-\left[\Delta_0+S(\textbf{k})+
\left[h_z-\xi_\downarrow(\textbf{k})\right]
\left[h_z+\xi_\uparrow(\textbf{k})\right]\right]-\left[\Delta^2+S(\textbf{k})+\left[h_z-\xi_\downarrow(\textbf{k})\right]\right]\Omega
+\left[h_z+\xi_\uparrow(\textbf{k})\right]\Omega^2(\textbf{k})+\Omega^3(\textbf{k})}
{2\Omega(\textbf{k})\left[\Omega(\textbf{k})-\omega(\textbf{k})\right]\left[\Omega(\textbf{k})+\omega(\textbf{k})\right]},$$
$$B_{11}(\textbf{k})=$$$$\frac{\left[-h_z+\xi_\downarrow(\textbf{k})\right]-\left[\Delta_0+S(\textbf{k})+
\left[h_z-\xi_\downarrow(\textbf{k})\right]
\left[h_z+\xi_\uparrow(\textbf{k})\right]\right]-\left[\Delta^2+S(\textbf{k})+\left[h_z-\xi_\downarrow(\textbf{k})\right]\right]\Omega
-\left[h_z+\xi_\uparrow(\textbf{k})\right]\Omega^2(\textbf{k})+\Omega^3(\textbf{k})}
{2\Omega(\textbf{k})\left[\Omega(\textbf{k})-\omega(\textbf{k})\right]\left[\Omega(\textbf{k})+\omega(\textbf{k})\right]},$$
$$C_{11}(\textbf{k})=$$
$$\frac{\left[-h_z+\xi_\downarrow(\textbf{k})\right]-\left[\Delta_0+S(\textbf{k})+
\left[h_z-\xi_\downarrow(\textbf{k})\right]
\left[h_z+\xi_\uparrow(\textbf{k})\right]\right]+\left[\Delta^2+S(\textbf{k})+
\left[h_z-\xi_\downarrow(\textbf{k})\right]\right]\omega
-\left[h_z+\xi_\uparrow(\textbf{k})\right]\omega^2(\textbf{k})-\omega^3(\textbf{k})}
{2\omega(\textbf{k})\left[\Omega(\textbf{k})-\omega(\textbf{k})\right]\left[\Omega(\textbf{k})+\omega(\textbf{k})\right]},$$
$$D_{11}(\textbf{k})=$$
$$\frac{\left[-h_z+\xi_\downarrow(\textbf{k})\right]-\left[\Delta_0+S(\textbf{k})+
\left[h_z-\xi_\downarrow(\textbf{k})\right]
\left[h_z+\xi_\uparrow(\textbf{k})\right]\right]+\left[\Delta^2+S(\textbf{k})+
\left[h_z-\xi_\downarrow(\textbf{k})\right]\right]\omega
+\left[h_z+\xi_\uparrow(\textbf{k})\right]\omega^2(\textbf{k})-\omega^3(\textbf{k})}
{2\omega(\textbf{k})\left[\Omega(\textbf{k})-\omega(\textbf{k})\right]\left[\Omega(\textbf{k})+\omega(\textbf{k})\right]}.$$
\end{widetext}


\begin{thebibliography}{999}
\bibitem{M} E. Majorana, Nuovo Cimento \textbf{5}, 171 (1937).
%
\bibitem{SOC1} Y. J. Lin, R. L.
Compton, K. Jimenez-Garcia, J. V. Porto, and I. B. Spielman, Nature \textbf{462},
628 (2009)
\bibitem{SOC2} J. Dalibard, F. Gerbier, G. Juzeliunas, and Patrik Ohberg, Rev. Mod. Phys.
\textbf{83}, 1523 (2011).
%
\bibitem{SOC3} J. D. Sau, Ra. Sensarma, S. Powell, I. B. Spielman, and S. Das Sarma, Phys.
Rev. B \textbf{83}, 140510(R) (2011).
%
\bibitem{SOC4} D. L. Campbell, G. Juzeliunas, and I. B. Spielman, Phys. Rev. A \textbf{84}, 025602
(2011).
%
\bibitem{SOC5} Y.-J. Lin, K. Jiménez-García, and I. B. Spielman, Nature
(London) \textbf{471}, 83 (2011).
%
\bibitem{SOC6} B. M. Anderson, G. Juzeliunas, I. B. Spielman, and V. M. Galitski, Phys.
Rev. Lett. \textbf{108}, 235301 (2012).
%
\bibitem{SOC7} P. Wang, Z.-Q. Yu, Z. Fu, J. Miao, L. Huang, S. Chai,
H. Zhai, and J. Zhang, Phys. Rev. Lett. \textbf{109}, 095301
(2012).
%
\bibitem{SOC8} L.W. Cheuk, A. T. Sommer, Z. Hadzibabic, T. Yefsah,
W. S. Bakr, and M.W. Zwierlein, Phys. Rev. Lett. \textbf{109},
095302 (2012).
%
\bibitem{SOC9} J.-Y. Zhang, S.-C. Ji, Z. Chen, L. Zhang, Z.-D. Du, B. Yan,
G.-S. Pan, B. Zhao, Y.-J. Deng, H. Zhai, S. Chen, and
J.-W. Pan, Phys. Rev. Lett. \textbf{109}, 115301 (2012).
%
\bibitem{SOC10} C. Qu, C. Hamner,M. Gong, C. Zhang, and P. Engels, Phys.
Rev. A \textbf{88}, 021604(R) (2013).
%
\bibitem{SOC11} R. A. Williams, M. C. Beeler, L. J. LeBlanc, K. Jiménez-
García, and I. B. Spielman, Phys. Rev. Lett. \textbf{111}, 095301
(2013).
%
\bibitem{Ham1} Yong Xu, C. Qu, M. Gong, and C. Zhang, Phys. Rev. A. \textbf{89},
013607 (2014).
%
\bibitem{Ham2} C. Qu, M. Gong, and C. Zhang, Phys. Rev. A. \textbf{89},
053618 (2014).
%
\bibitem{G1} J. R. Engelbrecht, M. Randeria, and C. A. R. S$\acute{a}$ de Melo, Phys. Rev. B
\textbf{55}, 15153 (1997).
%
\bibitem{Schwinger} J. Schwinger, Phys. Rev. \textbf{82}, 914
(1951).
%
\bibitem{Dyson} F. J. Dyson, Phys. Rev. \textbf{75}, 1736
(1949).
%
\bibitem{BetheS}  H. A. Bethe and E. E. Salpeter, Phys. Rev. \textbf{82}, 309
(1951); ibit.  \textbf{84}, 1232 (1951).
%
\bibitem{KMF} Z. Koinov, R. Mendoza and M. Fortes,  Phys. Rev. Lett. \textbf{106}
100402 (2011).
%
\bibitem{RM} R. Mendoza, M. Fortes, M. A. Sol\'{\i}s and Z. Koinov,   Phys. Rev. A. \textbf{88}
033606 (2013).
%
\bibitem{KPM} Z. Koinov, S. Pahl, and R. Mendoza,  J.Low Temp. Phys.  \textbf{173}, 231 (2015).
%
\bibitem{ZGK} Z. G. Koinov,  Advances in Condensed Matter Physics, vol. \textbf{2015}, 952852 (2015).
%
\bibitem{BR} L. Belkhir and M. Randeria, Phys. Rev. B \textbf{49}, 6829 (1994).
%
\bibitem{CKS} R. Combescot, M. Yu. Kagan, and S. Stringari, Phys. Rev. A
\textbf{74}, 042717 (2006).
%
\bibitem{GSOC1} Hui Hu, Lei Jiang, Xia-Ji Liu, and Han Pu, Phys. Rev. Lett. \textbf{107}, 195304 (2011).
%
\bibitem{GSOC2} K. Zhou and Z. Zhang, Phys. Rev. Lett. \textbf{108}, 025301 (2012).
%
\bibitem{GSOC3} K. Seo, Li Han, and C. A. R. S\'{a} de Melo, Phys. Rev. A \textbf{86}, 033601 (2012).
%
\bibitem{GSOC4} Lianyi He and Xu-Guang Huang, Phys. Rev. A \textbf{86}, 043618 (2012).
%
\bibitem{GSOC5} J. P. Vyasanakere. and V. B. Shenoy, Phys. Rev. A \textbf{86}, 053617 (2012).
%
\bibitem{GSOC6} L. He, Xu-Guang Huang, Annals of Physics \textbf{337}  163, (2013).
%
\bibitem{GSOC7}Yong Xu, Rui-Lin Chu, and C. Zhang, Phys. Rev. Lett. \textbf{112}, 136402 (2014).
%
\bibitem{GSOC8} Y. Xu, Rui-Lin Chu, and C. Zhang,  	Int. J. Mod. Phys. B  \textbf{29}, 1530001 (2015).
%
\bibitem{Jap} Y. Yunomae,  D.
Yamamoto, I. Danshita, N Yokoshi, and S Tsuchiya,  Phys. Rev. A \textbf{80}, 063627 (2009).
 %
\bibitem{RSOC} G. Liu, N. Hao, Shi-Liang Zhu, and W. M. Liu
Phys. Rev. A \textbf{86}, 013639 (2012).
%
\bibitem{FF} P. Fulde, and R. A. Ferrell,  Phys. Rev. \textbf{135}, A550 (1964).
%
\bibitem{Sarma} G. Sarma, J. Phys. Chem. \textbf{24}, 1029 (1963).
%

\bibitem{ZKexc} Z. Koinov,  Phys. Rev. B \textbf{72}, 085203 (2005).
%
\bibitem{GF} L. Jiang, Xia-Ji Liu, H. Hu, and H. Pu
Phys. Rev. A \textbf{84}, 063618 (2011)

\end{thebibliography}
\end{document}